\documentclass[a4paper]{jpconf}
\usepackage{graphicx}
\begin{document}
\newcommand\aj{{AJ}}%
\newcommand\araa{{ARA\&A}}%
\newcommand\apj{{ApJ}}%
\newcommand\apjl{{ApJ}}%
\newcommand\apjs{{ApJS}}%
\newcommand\ao{{Appl.~Opt.}}%
\newcommand\apss{{Ap\&SS}}%
\newcommand\aap{{A\&A}}%
\newcommand\aapr{{A\&A~Rev.}}%
\newcommand\aaps{{A\&AS}}%
\newcommand\azh{{AZh}}%
\newcommand\baas{{BAAS}}%
\newcommand\jrasc{{JRASC}}%
\newcommand\memras{{MmRAS}}%
\newcommand\mnras{{MNRAS}}%
\newcommand\pra{{Phys.~Rev.~A}}%
\newcommand\prb{{Phys.~Rev.~B}}%
\newcommand\prc{{Phys.~Rev.~C}}%
\newcommand\prd{{Phys.~Rev.~D}}%
\newcommand\pre{{Phys.~Rev.~E}}%
\newcommand\prl{{Phys.~Rev.~Lett.}}%
\newcommand\pasp{{PASP}}%
\newcommand\pasj{{PASJ}}%
\newcommand\qjras{{QJRAS}}%
\newcommand\skytel{{S\&T}}%
\newcommand\solphys{{Sol.~Phys.}}%
\newcommand\sovast{{Soviet~Ast.}}%
\newcommand\ssr{{Space~Sci.~Rev.}}%
\newcommand\zap{{ZAp}}%
\newcommand\nat{{Nature}}%
\newcommand\iaucirc{{IAU~Circ.}}%
\newcommand\aplett{{Astrophys.~Lett.}}%
\newcommand\apspr{{Astrophys.~Space~Phys.~Res.}}%
\newcommand\bain{{Bull.~Astron.~Inst.~Netherlands}}%
\newcommand\fcp{{Fund.~Cosmic~Phys.}}%
\newcommand\gca{{Geochim.~Cosmochim.~Acta}}%
\newcommand\grl{{Geophys.~Res.~Lett.}}%
\newcommand\jcp{{J.~Chem.~Phys.}}%
\newcommand\jgr{{J.~Geophys.~Res.}}%
\newcommand\jqsrt{{J.~Quant.~Spec.~Radiat.~Transf.}}%
\newcommand\memsai{{Mem.~Soc.~Astron.~Italiana}}%
\newcommand\nphysa{{Nucl.~Phys.~A}}%
\newcommand\physrep{{Phys.~Rep.}}%
\newcommand\physscr{{Phys.~Scr}}%
\newcommand\planss{{Planet.~Space~Sci.}}%
\newcommand\procspie{{Proc.~SPIE}}%
\newcommand\an{{Astron.~Nachr.}}%

\title{Time--distance inversions for horizontal and vertical flows on supergranular scales applied to MDI and HMI data}

\author{M \v{S}vanda$^{1,2}$, H Schunker$^3$ and R Burston$^3$}
\address{$^1$ Astronomical Institute, Academy of Sciences of the Czech Republic (v. v. i.), Fri\v{c}ova 298, CZ-25165 Ond\v{r}ejov, Czech Republic}
\address{$^2$ Charles University in Prague, Faculty of Mathematics and Physics, Astronomical Institute, V Hole\v{s}ovi\v{c}k\'ach 2, CZ-18000 Prague 8, Czech Republic}
\address{$^3$ Max-Planck-Institut f\"ur Sonnensystemforschung, Max-Planck-Strasse 2, D-37191~Katlenburg-Lindau, Germany}
\ead{michal@astronomie.cz}

\begin{abstract}
We study the possibility of consistent extension of MDI full-disc helioseismic campaigns with the growing data set of HMI observations. To do so, we down-sample and filter the HMI Dopplegrams so that the resulting spatial power spectrum is similar to the spatial power spectrum of MDI full-disc Dopplergrams. The set of co-spatial and co-temporal datacube pairs from both instruments containing no missing and no bad frames were processed using the same codes and inverted independently for all three components of the plasma flow in the near surface layers. The results from the two instruments are highly correlated, however systematically larger (by $\sim 20$\%) flow magnitudes are derived from HMI.  We comment that this may be an effect of the different formation depth of the Doppler signal from  the two instruments. 
\end{abstract}

\section{Motivations}

Understanding the details of the solar dynamo is inevitably linked to measurements of plasma flows inside the Sun. Methods of helioseismology \cite{2010ARAA..48..289G,2005LRSP....2....6G} can measure such flows, one of them being time--distance local helioseismology \cite{1993Natur.362..430D}. It is based on the measurements of travel times of seismic waves, which propagate through the solar interior and carry information about the inhomogeneities (flows in particular) affecting their travel time. By solving the inverse problem one can learn about these inhomogeneities. 

A proper assessment of the character and structure of solar flows requires a long term study. Recently, the Michelson Doppler Imager (MDI \cite{1995SoPh..162..129S}, on board of SOHO spacecraft), which covered 15 years of observations, was replaced by its successor Helioseismic and Magnetic Imager (HMI \cite{2012SoPh..275..229S}, on board of SDO spacecraft). Our study is thus driven by the need to fully exploit \emph{all} of the data. Here, we specifically ask the question: \emph{Can MDI and HMI data be possibly utilised together for inverse modelling for time--distance local helioseismology?} 

In local helioseismology, two methods are usually used to solve the inverse problem. The regularised least squares (RLS; in time--distance helioseismology used for the first time by \cite{1996ApJ...461L..55K}) minimisation seeks to find models of solar interior that minimise the difference between the predicted and measured travel times. The inverse problem must be solved separately for each set of travel times. On the contrary, optimally localised averaging (OLA; \cite{1968GeoJ...16..169B,1970RSPTA.266..123B,1992AA...262L..33P}) method is based on explicitly constructed spatially confined averaging kernels, while bounding the noise propagation at the same time. The inverse problem in the OLA approach must be solved only once and does not directly involve the measurements, except for the knowledge of statistical properties of the travel-time noise. Estimates for inverted quantities are then obtained by taking linear combination of travel times utilising a set of weights resulting from the inversion. When the measurements from various instruments are combined to form a homogeneous data set, the OLA is a natural approach to be used for inverse modelling from such data set. 

\begin{figure}[!t]
\includegraphics[width=0.45\textwidth]{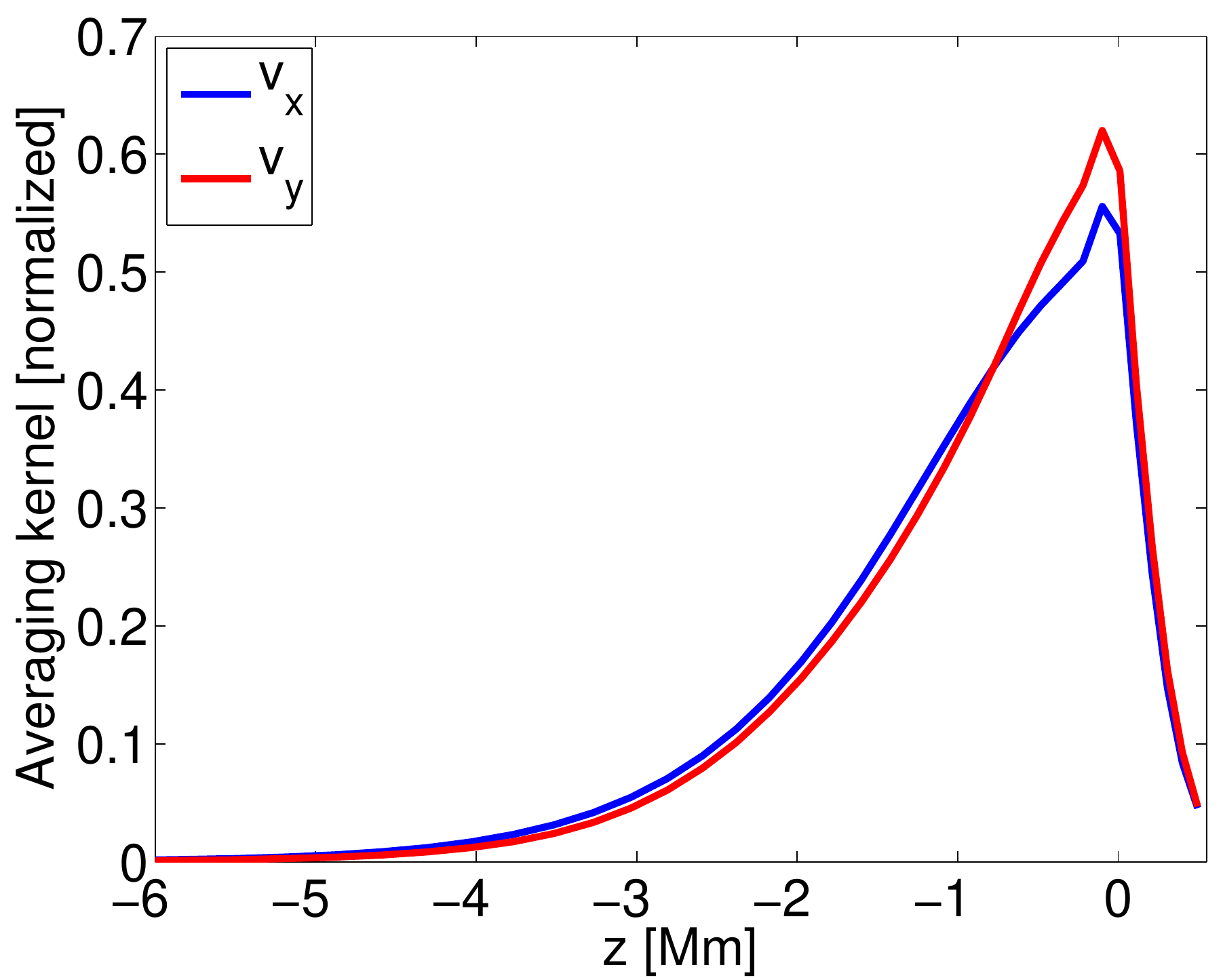}\hspace{2pc}%
\begin{minipage}[b]{18pc}\caption{\label{fig:akerns}Cuts through averaging kernels of inversion for horizontal $v_x$ and vertical $v_z$ flow in the middle point as a function of depth. Note that $z=0$ at an optical depth of $\tau_{500}=1$, negative $z$ values refer to solar interior and positive ones refer to atmosphere. }
\end{minipage}
\end{figure}

To test inverse modelling, we constructed a well-understood inversion for all three flow components. The inversion follows the Fourier-space based solution \cite{fastOLA} using the validated inversion code \cite{Svanda2011} implemented within the German Science Center for SDO at Max-Planck-Institut f\"ur Sonnensystemforschung in Katlenburg-Lindau, Germany. This code introduces additional terms in the cost function of the inverse problem, which allows the minimisation of the contributions from cross-talks. This property is crucial in order to be able independently invert also for a weak vertical flow. The inversion utilised a combination of wave sensitivity kernels computed in the Born approximation \cite{2007AN....328..228B} for the surface gravity wave ($f$) mode, assuming the application to the MDI full-disc Dopplergrams. The travel-time noise covariance matrix used in the inverse problem was measured from the data using $1/T$ fitting approach \cite{2004ApJ...614..472G}. 
The inversion combined sensitivity kernels for travel times for  three point-to-annulus and point-to-quadrant geometries (making the measurements sensitive to divergence of the flow and vertical velocity, and to waves travelling in west--east and north--south directions) with radii of the annuli from 7.3 to 29.2~Mm. The inversion targets only near sub-surface layers (see the averaging kernels in Fig.~\ref{fig:akerns}).

\section{Data}
To do a proper comparison between the two instruments, we required the consistent analysis of the data, which are both co-spatial and co-temporal. MDI delivered data suitable for local helioseismology only two months a year (during Dynamics Campaigns), with the last one (June and July 2011) overlapping with routine operations of HMI. In this two-month period we prepared 64 consecutive pairs (one for MDI, one for HMI) of 12-hour long Dopplergram datacubes capturing the centre of the solar disc using the mapping and tracking pipeline at the GDC-SDO. Additionally, we did not want to have results affected by missing or bad frames in datacubes, thus we required a 100\% duty cycle. A down-selection was done on a frame-by-frame basis manually. Only 4 (!) out of 64 co-spatial and co-temporal datacube pairs fulfilled our requirements. 

\begin{figure}[!t]
\includegraphics[width=0.99\textwidth]{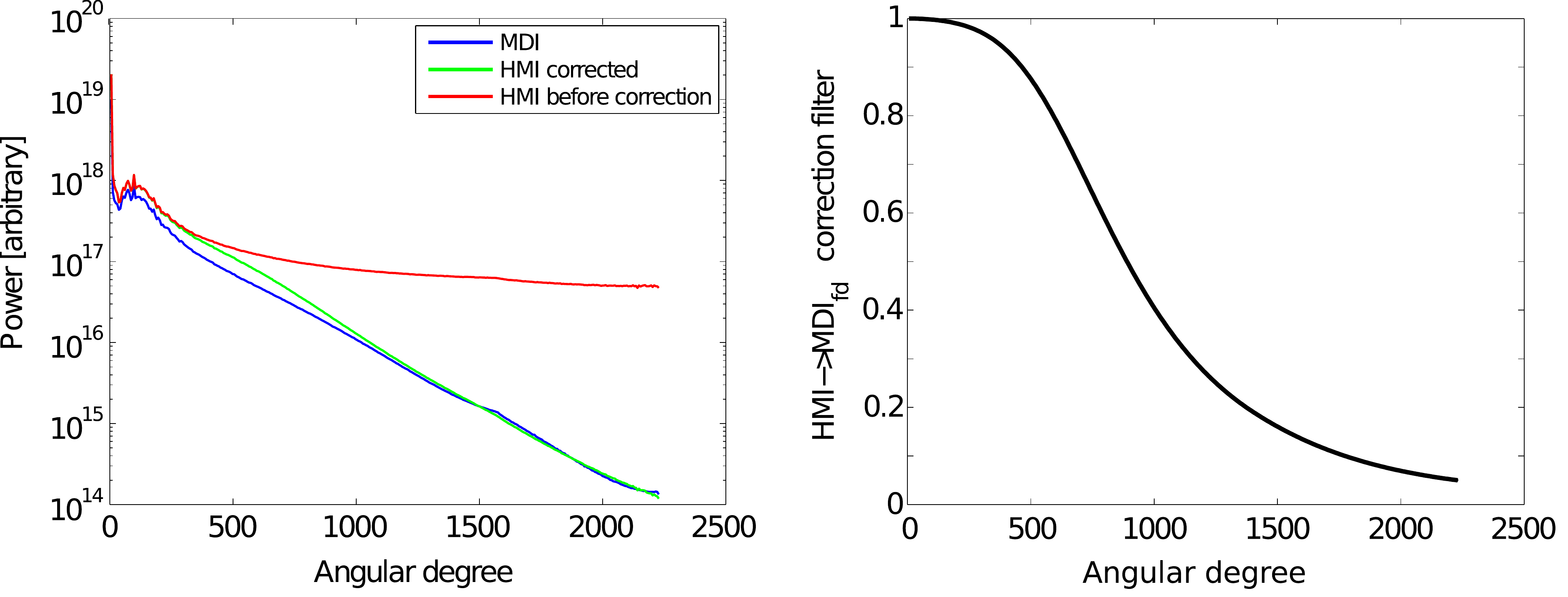}
\caption{Left: Spatial power spectra of MDI$_{\rm fd}$ Dopplergrams compared to HMI Dopplergrams down-sampled to MDI$_{\rm fd}$ resolution before and after correction. Right: The wave-number filter designed for the correction. }
\label{fig:correction}
\end{figure}

The sensitivity kernels used in the inverse problem described above were designed to be consistent with MDI full-disc (MDI$_{\rm fd}$ hereafter) Dopplergrams. HMI is a higher-resolution instrument, thus we needed to make sure that HMI data looked like MDI$_{\rm fd}$ ones. This was achieved by a Postel-projection of both MDI and HMI Dopplergrams with the same map scale (corresponding to MDI$_{\rm fd}$ pixel size) using bicubic interpolation. HMI Dopplergrams were thus down-sampled to MDI$_{\rm fd}$ resolution. Even in such data the contribution from large wave-number was much higher than in the case of true MDI$_{\rm fd}$ data (see Fig.~\ref{fig:correction} left) and we wanted to correct for it. This may be done by an application of a filter in the wave-number space (Fig.~\ref{fig:correction} right), designed so that the spatial power spectra of both instruments decay similarly in the large wave-number region. This filter approximates the correction to the different modulation transfer functions (MTF) of the two instruments. Note that despite the filtering, spatial power spectrum of corrected HMI Dopplergrams still have an excess in power at angular degrees $l$ of 50 to 300. We will return to this issue in the next section.

Apart from the power-spectrum correction, the rest of the analysis (filtering, travel-time measurements, application of inversion weights) was identical for both MDI and degraded HMI datacubes. 

\section{Comparison of helioseismic observables}

\begin{figure}[!t]
\includegraphics[width=0.99\textwidth]{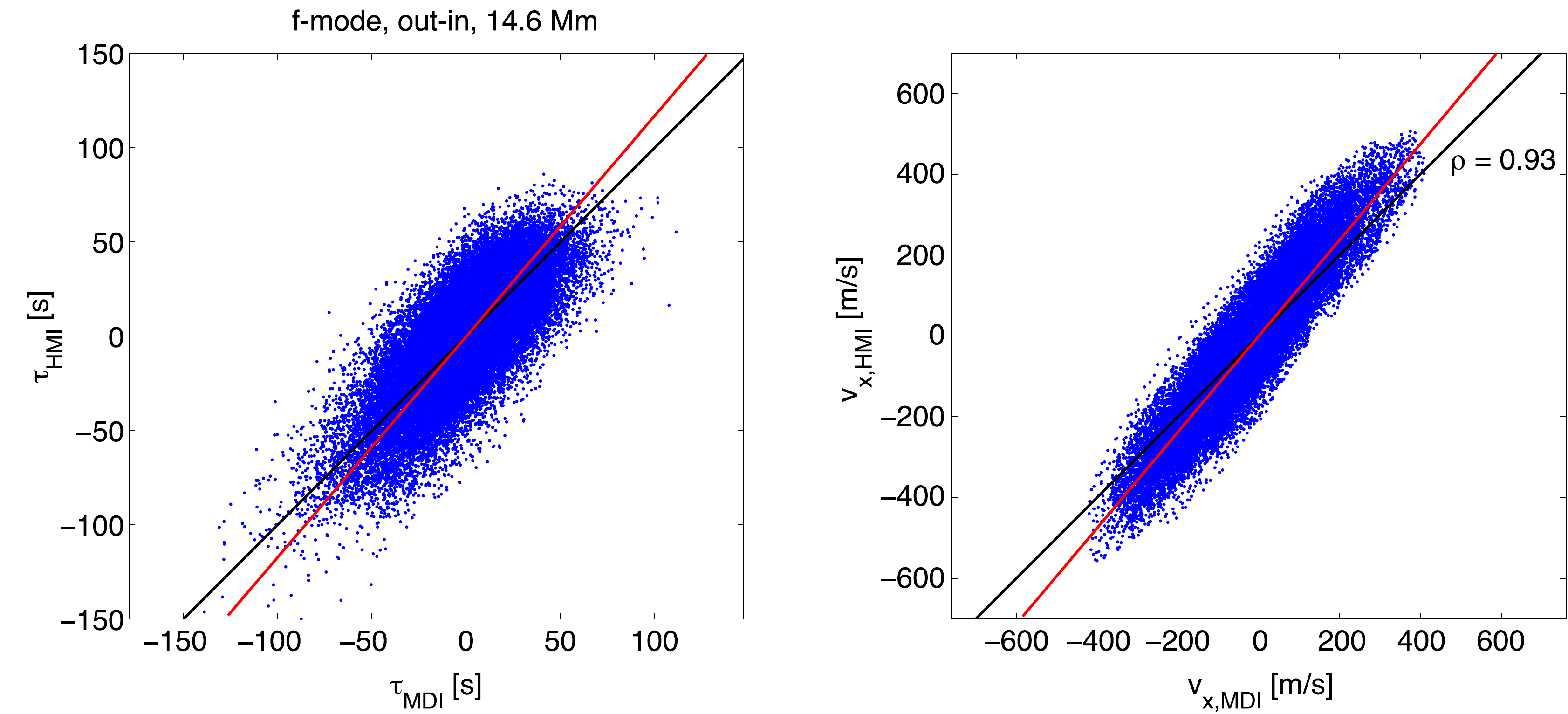}
\caption{Left: Scatter plot of the travel times measured for the $f$-mode with point-to-annulus geometry with distance 10~px from MDI$_{\rm fd}$ (horizontal axis) and degraded HMI (vertical axis). Right: Equivalent scatter plot to compare the inversion estimate of $v_x$ with the correlation coefficient indicated. In both cases, the red solid line represents the best least-squares fit of the points and the black solid line has a unity slope to increase the visibility of the systematically larger values derived from HMI Dopplergrams. }
\label{fig:crossplots}
\end{figure}

Consistently with the setup of the inversion, we measured wave travel times from $f$-mode-pass filtered Dopplergram datacubes with point-to-annulus and point-to-quadrants geometries with distances 7.3 to 29.2~Mm. Travel-time maps measured from corresponding MDI and HMI datacubes are highly correlated (correlation coefficients of 0.8 or higher), but the travel times measured from HMI datacubes are systematically  larger by 20\% (e.g., Fig.~\ref{fig:crossplots} left). This statement holds for all pairs of travel-time maps. The excess is mostly prominent on supergranular and similar scales (angular degree between 50 and 300) and it seems to be a consequence of the excess power discussed above (see Fig.~\ref{fig:correction} left). 

Since the estimate of the inverted flows is just a linear combination of travel times and the inversion weights are the same for both MDI and HMI measurements, the magnitudes of flow estimates are also systematically larger, when the flow is inverted from the HMI datacubes than from the MDI datacubes (see example in Fig.~\ref{fig:crossplots} right). Apart from this systematic effect (there is an excess of 20\% for horizontal components and around 30\% for vertical component) the flows derived from MDI and HMI measurements are almost identical (correlation coefficient larger than $0.9$ for horizontal components and around $0.8$ for the vertical component). The lower correlation in the case of vertical component is probably caused by lower signal-to-noise ratio of the vertical flow estimate. 

\begin{figure}[!t]
\includegraphics[width=0.99\textwidth]{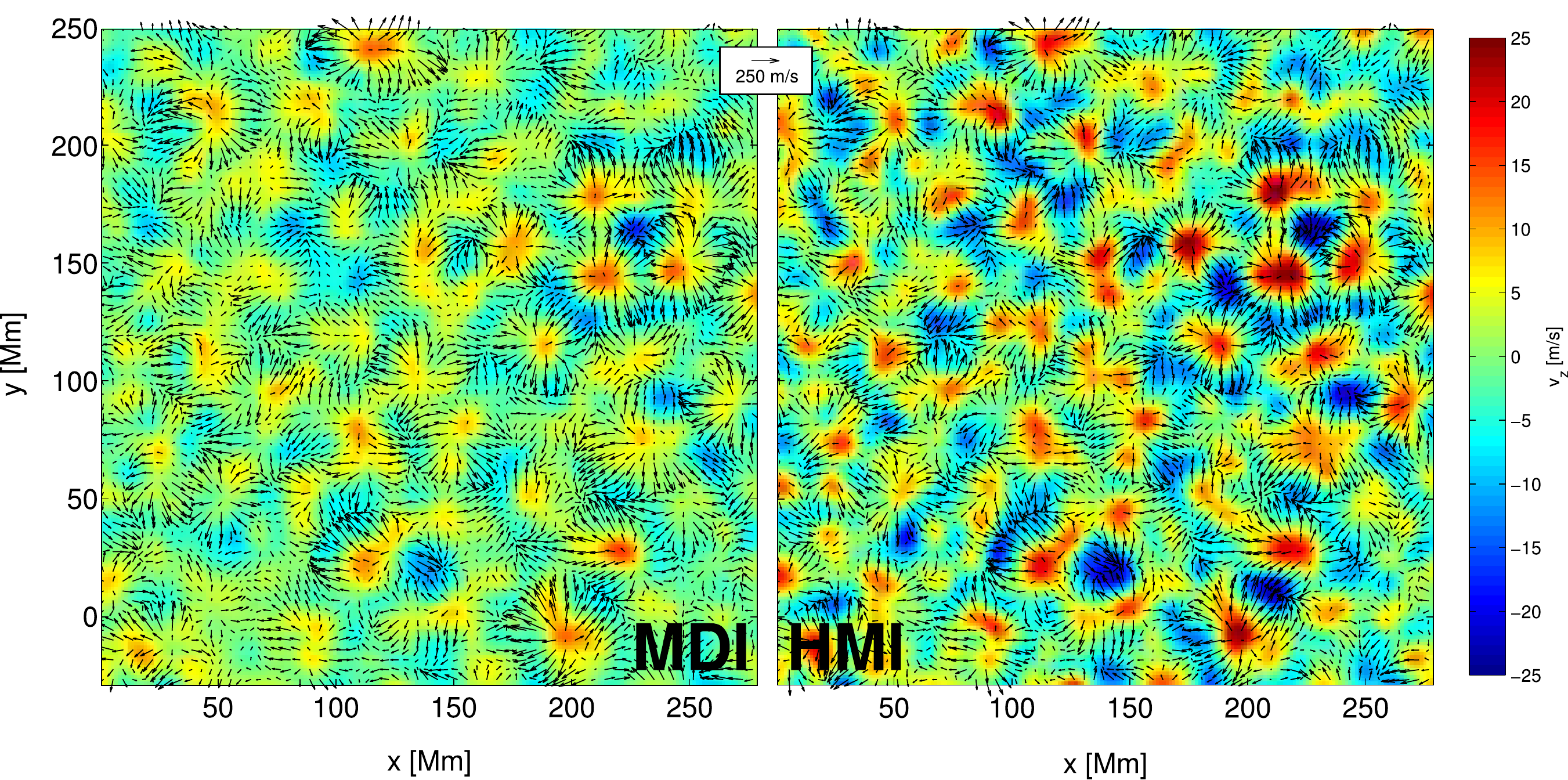}
\caption{Comparison of full-vector flow maps in the same region on the Sun, inverted from MDI (left) and HMI (right) Dopplergrams. Horizontal components in the plane of solar surface are indicated by arrows, the vertical component by colours. The level of random error in the flow estimates is 49~m/s for horizontal flow components and 4~m/s for the vertical one. }
\label{fig:fullflows}
\end{figure}

An example of full-vector flow maps for HMI and MDI is displayed in Fig.~\ref{fig:fullflows}. One can see that except for the systematically larger magnitudes in the HMI flow map, both maps are practically identical. In these maps, supergranular cells (rosette-like features in the horizontal components) are dominant. Also note that the upflows (red) in the vertical component in most cases correspond to outflows in the horizontal component. The correlation between the vertical component and the horizontal divergence of the horizontal flow is higher than 0.8. This is one indication that the inversion for the vertical flow is reasonable. 

\section{Final remarks}
We showed that with some drawbacks, HMI and MDI helioseismic datasets may be analysed together. Long-term studies of flows in the solar convection zone employing combined datasets are therefore possible. However, the noticed bias must be understood and corrected for. 

It is likely that the different flow magnitudes are an effect of the different spectral line used in the two instruments. The formation depth of HMI Doppler signal measured in Fe\,I (617.3~nm) line is some 25~km deeper in the photosphere than the formation depth of the MDI Doppler signal measured in Ni\,I (676.8~mm) line \cite{2011SoPh..271...27F}. Thus the convection (especially supergranules) is better pronounced when observed in the Fe\,I line, hence the power spectrum is expected to have more power on scales where convection is dominant, including the supergranular scales around angular degree of 120. That is the case and it can be clearly seen in Fig.~\ref{fig:correction}.  Increased sensitivity to the convection affects the sensitivity kernels, which slightly differ for the two spectral lines (however, we did not compute the proper sensitivity kernels for Fe\,I line, so we did not model the effect properly). As a consequence, we expect that when the weights from the inversion performed with Ni\,I-line sensitivity kernels are applied to travel-time maps measured from Fe\,I line, the true inversion averaging kernel is slightly different, with the peak sensitivity located probably a little deeper (the averaging kernels for the Ni\,I-line inversion are plotted in Fig.~\ref{fig:akerns}). If we assume that the magnitude of the plasma flow increases with height in the last 1 Mm of the convection zone and then drops gradually around $\tau_{500}=1$ level, the averaging kernel with a deeper peak averages the flow with slightly higher amplitudes and therefore the resulting flow estimate will also have larger magnitude. That is observed as a bias when comparing the flow estimates from the two instruments. The higher bias in the vertical flow ($\sim 30$\%) than in the horizontal components ($\sim 20$\%) might indicate a steeper gradient of the vertical flow below the surface.

Other effects may also be responsible for the excess power in the spectrum, such as a more complicated instrumental MTF shape which was not corrected for. After the bias is understood, it would be possible to correct for it, for example by application of an additional non-monotonic wave-number filter. 

This study is  the first to compare inverse modelling using the OLA technique for MDI and HMI data. Similar direct comparisons of flow and sound speed inversions was performed on early HMI data (J~Zhao, private communication) using the RLS technique, but it showed larger discrepancies not only in magnitude, but also in structure of the flows than presented here. This higlights the necessity to first test and understand the systematics due to both the instrument and the analysis technique.

\ack 
M\v{S} acknowledges the support of the Czech Science Foundation (grant P209/12/P568). This work utilised the resources and helioseismic products 
dispatched within the German Science Center at MPS in Katlenburg-Lindau, Germany, which is supported by German 
Aerospace Center (DLR). The data were kindly provided by the HMI consortium. The HMI project is supported by NASA contract NAS5-02139. Tato pr\'ace vznikla s podporou na dlouhodob\'y koncep\v{c}n\'\i{} rozvoj v\'yzkumn\'e organizace (RVO:67985815).
\\

 \bibliographystyle{iopart-num}
 \bibliography{inversions}

\end{document}